\documentclass[
    nofootinbib, 
    amsmath, 
    amssymb, 
    aps, 
    pra, 
    twocolumn,
    superscriptaddress,
    10pt
]{revtex4-2}

\usepackage{physics, nicefrac, stmaryrd}
\usepackage{graphicx}
\usepackage[table]{xcolor}
    \definecolor{myblue}{RGB}{3,70,143}
\usepackage[colorlinks,citecolor=myblue,linkcolor=myblue,urlcolor=myblue]{hyperref}
\usepackage{dsfont}

\usepackage[capitalise]{cleveref} 

\newcommand{\id}{\ensuremath{\mathds{1}}}
\newcommand{\eps}{\ensuremath{\varepsilon}}
\DeclareMathOperator*{\argmax}{arg\,max}

\begin{document}

\title{Optimal Decoding of Small Codes by Density Matrix Propagation}

\author{Anthony Benois}
\affiliation{Université Paris–Saclay, CEA, CNRS, Institut de Physique Théorique, 91191 Gif-sur-Yvette, France}

\author{Pierre Cussenot}
\affiliation{Université Paris–Saclay, CEA, CNRS, Institut de Physique Théorique, 91191 Gif-sur-Yvette, France}
\affiliation{Direction Générale de l’Armement, 75015 Paris, France}

\author{Grégoire Misguich}
\affiliation{Université Paris–Saclay, CEA, CNRS, Institut de Physique Théorique, 91191 Gif-sur-Yvette, France}

\author{Nicolas Sangouard}
\affiliation{Université Paris–Saclay, CEA, CNRS, Institut de Physique Théorique, 91191 Gif-sur-Yvette, France}

\author{Kiara Hansenne}
\affiliation{Université Paris–Saclay, CEA, CNRS, Institut de Physique Théorique, 91191 Gif-sur-Yvette, France}

\date{\today}

\begin{abstract}
Accurate and efficient decoding is a crucial component for achieving fault-tolerant quantum computing.
Realistic circuit-level noise introduces temporal correlations and degeneracy, making optimal (maximum-likelihood) decoding computationally intractable in general.
As a result, practical decoders rely on heuristic approximations, and it is generally difficult to quantify how suboptimal they are, as this strongly depends on the code and noise model considered.
In this work, we study the accuracy of practical decoding algorithms under circuit-level noise by comparing them against a maximum likelihood decoding benchmark. 
Our approach propagates the density matrix through the full memory experiment and computes the optimal decoding decision for each syndrome history. 
We introduce pruning techniques with rigorous bounds, allowing us to access larger numbers of syndrome-extraction rounds.
We apply this framework to small instances of the repetition code and a cellular automaton code, and benchmark minimum-weight perfect matching (MWPM), belief propagation with ordered statistics decoding (BP+OSD), Tesseract, and Planar decoders against optimal decoding.
While standard decoders remain close to optimal for the repetition code, we find significant deviations for the cellular automaton code, with BP+OSD deteriorating already in experimentally relevant noise regimes.
Moreover, the pruning method developed here highlights that, at low physical error rates, only a narrow fraction of syndrome histories contributes significantly to the logical error rate.
\end{abstract}

\maketitle

\section{Introduction}
    Quantum computing promises to solve problems inaccessible to classical computers~\cite{shor1994algorithms,lloyd1996universal,aspuru2005simulated,preskill2025beyond,gouzien2023performance,zhao2026exponential}.
    However, quantum systems are highly sensitive to noise and decoherence.
    Practical quantum computers therefore need \emph{fault-tolerance}, which relies on \emph{quantum error correction} (QEC) \cite{terhal2015quantum,campbell2017roads}.
    In QEC, quantum information is redundantly encoded across many physical qubits such that, below a threshold physical error rate, logical errors become exponentially suppressed with the code distance~\cite{aharonov1997fault,knill1998resilient,terhal2015quantum}.
    Recent experiments have demonstrated this error suppression on some hardware platforms~\cite{google2025quantum,he2025experimental}, while several others have reported results approaching this regime~\cite{putterman2025hardware,bluvstein2026fault,atom2026quantum}, yet making QEC truly practical remains one of the central challenges in the field.

    While many QEC codes with desirable experimental properties are known~\cite{bombin2006topological,fowler2012surface,breuckmann2021quantum,panteleev2022asymptotically,bravyi2024high}, their practical usefulness critically depends on decoding, namely the process of interpreting noisy measurement outcomes in order to apply the most effective corrective operation.
    Decoding is a major bottleneck for fault-tolerant quantum computing: without fast and accurate decoders, errors cannot be sufficiently suppressed~\cite{terhal2015quantum,demarti2024decoding}.

    The performance of a decoder strongly depends on both the considered code and the underlying noise model.
    A decoder may perform near-optimally for one code while failing badly for another~\cite{roffe2020decoding}, which becomes even more pronounced under realistic circuit-level noise models, where syndrome extraction itself is noisy~\cite{dennis2002topological, fowler2012surface}.
    Optimal decoding, known as \emph{maximum-likelihood} (ML) decoding, must then account for large equivalence classes of error histories and is $\#$P-complete~\cite{poulin2008iterative,iyer2015hardness}.
    As a result, practical decoding relies on heuristic approximations.

    The standard approach to assess the accuracy of a decoder is based on Monte Carlo sampling of noisy trajectories and estimating the resulting logical error rate~\cite{fowler2012surface, gidney2021stim}.
    For stabiliser circuits and Pauli noise, such simulations can be performed efficiently~\cite{gottesman1998heisenberg, aaronson2004improved, anders2006fast} and allow one to compare heuristic decoders against one another.
    While this approach provides a practical benchmark, it does not directly quantify the gap to optimal decoding~\cite{iyer2015hardness,roffe2020decoding}. 
    Consequently, it is often unclear whether observed logical error rates are limited by the code and noise model themselves, or by the decoder's performance. 
    This ambiguity motivates the need for methods that can assess decoder performance relative to the ML limit for any QEC code.

    To address this issue, we present a framework based on the propagation of density matrices through noisy QEC circuits and compute ML logical error rates under circuit-level noise.
    The method recursively explores the syndrome tree associated with the memory experiment and evaluates the contribution of each syndrome history to the logical error rate.
    
    However, exact syndrome-tree exploration scales exponentially with the number of syndrome-extraction rounds. 
    To mitigate this, we introduce a pruning strategy with rigorous upper and lower bounds on the logical error rate.
    This pruning highlights that, at low physical error rates, only a small fraction of syndrome histories contributes significantly to the logical error rate, while the vast majority of syndrome histories are either extremely unlikely or unlikely to produce logical failures.
    
    Moreover, exhaustive syndrome exploration provides detailed information about syndrome distribution and contribution to the logical error rate. 
    In particular, it reveals how the probability and impact of individual syndrome histories evolve over successive rounds of syndrome extraction. 
    A syndrome that initially contributes significantly to the logical error rate may become considerably unlikely at later rounds, making its contribution ultimately negligible. 

    As an application, we benchmark several state-of-the-art decoders, namely MWPM \cite{higgott2025sparse}, BP+OSD \cite{roffe2020decoding}, Tesseract \cite{beni2025tesseract}, and Planar \cite{feng2025planar}, against this exact ML reference for the repetition code \cite{google2021exponential} and for a family of cellular automaton codes \cite{chowdhury1994design, ruiz2025ldpc, ruiz2025private}.
    We identify regimes where heuristic decoders depart significantly from ML decoding already at distances as small as five.
    While the method is restricted to relatively small codes and small number of rounds, exact benchmarks are useful for decoder validation, assessing practical decoders, and the analysis of proof-of-principle quantum error-correction experiments.

\section{Stabiliser codes and decoding}
    We begin with a brief introduction to stabiliser quantum error correction \cite{nielsen2010quantum}.
    This section reviews standard material, included for completeness and to fix notation.

    Quantum error correction aims to protect quantum information against noise by encoding it redundantly across multiple physical qubits using an $\llbracket n, k, d \rrbracket$ code.
    Concretely, a $k$-qubit logical state $\ket{\psi}$ is embedded into an $n$-qubit system, such that it lies within a $2^{k}$-dimensional subspace $\mathcal{C}$ of the full Hilbert space. 
    This subspace is referred to as the \emph{code space}.
    
    In the case of \emph{stabiliser codes}, the code space $\mathcal{C}$ is defined as the common $+1$-eigenspace of an abelian subgroup $\mathcal{S} \subset \mathcal{P}_n$ of the $n$-qubit Pauli group, such that $- \id \notin \mathcal{S}$, known as the \emph{stabiliser}.
    The stabiliser contains $2^{n-k}$ elements and can be generated by a set of $n-k$ independent commuting Pauli operators, called the stabiliser \emph{generators}.
    The code is therefore fully specified by these generators.
    
    After encoding, the $n$-qubit state is subject to noise.
    Without loss of generality, any resulting error $E$ can be represented as a Pauli operator, $E \in \mathcal{P}_n$.
    This follows from the fact that arbitrary error channels can be decomposed in the Pauli basis, and stabiliser measurements project them onto discrete Pauli errors.

    The action of an error can be understood in terms of its effect on the code space.
    If $E$ preserves $\mathcal{C}$, it acts as a \emph{logical operator} and is therefore \emph{undetected}.
    Otherwise, it maps states outside the code space.
    In that case, there exists at least one generator $S \in \mathcal{S}$ such that $S E\ket{\psi} = -E\ket{\psi}$. 
    Measuring the $n-k$ generators identifies the subspace in which $E\ket{\psi}$ lies.

    The outcomes of these measurements, assumed perfect for now, define the {syndrome}, a binary string $s \in \{0,1\}^{n-k}$.
    The syndrome $s$ is then processed by a \textit{decoder}, a classical algorithm that outputs a recovery operator $\mathcal{R} = \mathcal{D}(s)$.
    Formally, the decoder can be viewed as a map $\mathcal{D}: \{0,1\}^{n-k} \to \mathcal{P}_n$, taking syndromes to recovery operators.
    The goal of the decoder is to choose $\mathcal{R}$ such that it reverses the effect of the error, ideally satisfying $\mathcal{R} E \ket{\psi} = \ket{\psi}$. 
    However, the syndrome does not uniquely specify the error: many distinct errors can lead to the same syndrome.
    As a result, decoding is inherently ambiguous, and the recovery may fail with non-zero probability, leading to a \emph{logical error}.
    
    The fundamental parameter governing error correction is the \emph{distance} $d$ of the code, defined as the minimum weight of a Pauli operator that preserves the code space but acts non-trivially on the logical information.\footnote{The weight of a multi-qubit Pauli operator is the number of qubits on which it acts non-trivially.}
    Equivalently, it is the minimum weight of an undetectable error.
    Errors of weight strictly less than $d$ are therefore always detected. 
    Moreover, any error of weight strictly less than $\nicefrac{d}{2}$ is guaranteed to be corrected, in the sense that an ML decoder can identify a recovery operator $\mathcal{R}$ such that $\mathcal{R}E$ acts trivially on the code space.

    Most practical quantum decoders, such as MWPM~\cite{higgott2025sparse}, MWPF~\cite{wu2025minimum}, and Tesseract \cite{beni2025tesseract}, attempt to solve a \emph{maximum error probability} (MEP) problem \cite{iyer2015hardness}.
    Given a measured syndrome $s$, MEP decoding identifies the most likely error $E$ that could have produced it,
    \begin{equation}
        \hat{E} = \argmax_{E \in \mathcal{E}(s)} \mathbb{P}(E),
    \end{equation}
    where $\mathcal{E}(s)$ are all errors consistent with syndrome $s$.
    However, this strategy can be suboptimal due to the presence of \emph{degeneracy} in some stabiliser codes.
    
    A stabiliser code is said to be degenerate when two distinct errors $E_1$ and $E_2$ have the same effect on all codewords. 
    This occurs whenever they differ by a stabiliser element $S$, that is, when $E_2 = E_1 S$.
    Since $S$ stabilises the code space, 
    \begin{equation}
        E_2 \ket{\psi} = E_1 S \ket{\psi} = E_1 \ket{\psi}, 
    \end{equation}
    for all $\ket{\psi} \in \mathcal{C}$.
    The two errors are therefore physically indistinguishable and belong to the same \emph{equivalence class}.
    In such cases, a decoder does not need to identify the exact error, but only a representative of the corresponding class, as a single recovery operation suffices for all its elements.
    
    On the one hand, degeneracy increases the error-correcting capability of the code, as many physical errors become effectively equivalent.
    On the other hand, it means that MEP decoding is suboptimal: a MEP decoder picks the single most likely error while potentially ignoring a more likely equivalence class.
    
    Alternatively, {maximum-likelihood} decoders identify the most likely equivalence class rather than the most likely error~\cite{cao2026maximum,iyer2015hardness}. 
    Examples include Planar for the repetition code~\cite{feng2025planar} and tensor-network-based decoders for more general statistical-mechanical mappings~\cite{chubb2021general,bravyi2014efficient,piveteau2024tensor}.
    Equivalence classes are best described using the \emph{TLS decomposition}~\cite{iyer2015hardness}.
    For a stabiliser code with stabiliser $\mathcal{S}$, any Pauli error $E \in \mathcal{P}_n$ can be decomposed as
    \begin{equation}
        E = T L S,
        \label{eq:TLS_decomposition}
    \end{equation}
    up to a phase, where $S \in \mathcal{S}$ is a stabiliser element, $L \in \mathcal{P}_N$ is a logical operator, and $T\in \mathcal{P}_N$ is a representative of the error determined by the syndrome.
    Once a convention for choosing $T$ and a set of logical operators $L$ are fixed, this decomposition is unique \cite{fuentes2021degeneracy}. 

    This decomposition separates the error into three conceptually distinct parts: the stabiliser component $S$, which is irrelevant for the decoder as it acts trivially on codewords; the logical component $L$, which determines whether a logical error has occurred; and the pure error $T$, which captures the syndrome information.

    For a fixed syndrome $s$, the pure error $T$ is already determined. 
    The decoding problem therefore reduces to identifying the logical component $L$. 
    Indeed, after applying a recovery operation that restores the state to the code space, the remaining action of the physical error is entirely described by a logical operator. 
    A logical error occurs whenever this operator acts non-trivially on the encoded state. 
    Consequently, the goal of ML decoding is not to determine the exact physical error, but rather to identify the most likely logical operator compatible with the measured syndrome.

    Errors that share the same syndrome and logical component should therefore be equivalent from the decoder's perspective.
    An equivalence class can be labelled by $\mathcal{E}(s, L)$, and consists of all errors that share the same syndrome $s$ and logical component $L$, i.e., $\mathcal{E}(s, L) = \{E \mid E = T L S,  S \in \mathcal{S}\}$, where $T$ is the representative error of $s$.

    An ML decoder therefore selects the logical operator $L$ (or equivalently the equivalence class $\mathcal{E}(s,L)$) with the largest total probability~\cite{iyer2015hardness},
    \begin{equation} \label{eq:pbm_ml_decoding}
        \hat{\mathcal{E}} = \argmax_{\mathcal{E}(s, L)} \sum_{E \in \mathcal{E}(s, L)} \mathbb{P}\left(E \right), 
    \end{equation}
    and the recovery can be taken as any error in $\hat{\mathcal{E}}$.
    By construction, ML decoding identifies the correction that maximises the probability of successfully restoring the encoded state, and is therefore optimal.
    
    The optimality comes at a significant computational cost.
    MEP decoding is already NP-complete, while ML decoding is $\#$P-complete, as it requires summing over exponentially many errors within each equivalence class~\cite{iyer2015hardness}.
    Evaluating expressions such as \cref{eq:pbm_ml_decoding} thus amounts to a counting task.
    In practice, however, a decoder must output recovery operations at least as fast as syndromes are generated to avoid the backlog problem~\cite{terhal2015quantum}. 
    This requirement usually necessitates heuristic approximations, creating a trade-off between computational efficiency and decoding accuracy.
    
    The discussion so far assumes perfect syndrome extraction.
    In realistic settings, however, syndrome measurements are themselves noisy, and must be repeated to distinguish data errors from measurement errors.
    Each repetition defines a round, and the full syndrome (the syndrome history) is obtained by combining measurement outcomes across different rounds.

    This setting introduces an additional source of ambiguity in the decoding task, which we refer to as \emph{temporal degeneracy}.
    Unlike the degeneracy discussed above, which is a property of the code, temporal degeneracy arises from the noise model.
    It corresponds to the fact that multiple error histories can produce the same syndrome history and have the same action on the code space.
    
    Although temporal degeneracy is not always classified as genuine degeneracy in the literature, it leads to the same fundamental consequence: MEP decoding becomes suboptimal, and ML decoding must again account for entire equivalence classes of error histories.
    
    This can be understood within essentially the same decoding framework. 
    One considers an error pattern $E_1$, which can be represented as a set of faulty edges in the matching graph\footnote{The matching graph is a weighted graph whose vertices correspond to syndrome defects (or detection events), while edges represent possible error events connecting pairs of defects \cite{dennis2002topological, fowler2012surface}.}~\cite{higgott2025sparse}, together with a cycle $C$, defined as a closed loop of such edges that acts trivially on the codewords.
    A second error pattern can then be constructed as $E_2 = E_1 \oplus C$. 
    The two error patterns belong to the same ML equivalence class.
    When $C$ is purely space-like, it corresponds to a stabiliser and the resulting degeneracy is the conventional one encountered in stabiliser codes. 
    More generally, cycles with a temporal component give rise to temporal degeneracy. 
    As a consequence, even codes that are not intrinsically degenerate can exhibit degeneracy in the presence of noisy syndrome measurements.
    
    Some works treat these temporal degeneracies on the same footing as conventional degeneracies by promoting the corresponding cycles to formal stabilisers~\cite{feng2025planar,chubb2021statistical}. 
    This idea is motivated by the fact that noisy syndrome measurements can be incorporated into an extended stabiliser code. 
    This is done by introducing an ancilla qubit for each generator and each measurement round, and replacing each measured stabiliser $S$ with $S \mapsto S \otimes \widetilde{Z}$, where $\widetilde{Z}$ acts on the corresponding ancilla qubit~\cite{chubb2021statistical}. 
    Bit-flip errors on the ancilla then model measurement errors. 
    In this extended formulation, temporal degeneracy appears as ordinary stabiliser degeneracy.
    
    In this work, we benchmark heuristic decoders against ML decoding on effectively classical $[n, k, d]$ codes that are not conventionally degenerate but nevertheless exhibit temporal degeneracy.
    While the approach remains limited to moderate code sizes, we benchmark codes up to $[10,3,5]$.

\section{Density matrix propagation decoding}

    We assess decoder accuracy through simulations of \emph{memory experiments}.
    Memory experiments are a key benchmark for quantum error correction that aims to estimate the \emph{logical error rate} of encoded qubits. 
    
    A memory experiment begins by encoding a known logical state on $n$ physical qubits, typically the logical zero state $\ket{0_L}$ or the logical plus state $\ket{+_L}$, depending on whether logical bit-flips or phase-flips error rates are estimated.
    Without loss of generality, we present here the case of $\ket{0_L}$.
    The system is then subject to $r$ rounds of noisy syndrome extraction, during which the generators of the stabiliser are measured using ancillary qubits.
    In practice, the number of rounds is chosen to scale with the code distance, $r = \mathcal{O}(d)$~\cite{zhou2025low}.

    After the final round, the logical operator associated with the initial encoded state is measured, for example $Z_L$ if the encoded state is $\ket{0_L}$.
    Since this logical measurement may itself be faulty, the remaining data qubits are also measured to reconstruct the final stabiliser outcomes.
    Altogether, this produces a syndrome history $s \in \{0,1\}^{n_s}$ of length $n_s = (r+1) (n-k)$, which is given to a decoder that outputs a recovery operation $\mathcal{R}_s$.
    One can then explicitly verify whether applying $\mathcal{R}_s$ restores the initially encoded state $\ket{0_L}$.
    Note that for certain classes of codes, symmetries help to reduce the number of stabiliser values to reconstruct from the final data qubit measurements.
    This is for instance the case for the surface code, where only $Z$-type stabilisers are reconstructed.
    
    If the initially encoded state is not recovered, the experiment is said to result in a \emph{logical error}.
    The {logical error rate} $\eps_L \in [0,1]$ is defined as the probability of such a failure. 
    It depends both on the underlying physical noise model and on the decoding strategy.

    In memory experiments, the logical bit-flip error rate and the logical phase-flip error rate are typically estimated separately through the encoding of the corresponding basis state, and then averaged \cite{google2025quantum, putterman2025hardware, bluvstein2026fault}. 

    Next, we introduce the \emph{density matrix propagation} (DMP) decoding method. 
    It is based on the density matrix formalism, where all possible error trajectories are tracked and naturally grouped into equivalence classes. 
    Consequently, this decoding method belongs to the class of ML decoders.
    Moreover, the DMP method is not restricted to Clifford noise.
    We then present how it can be used to estimate logical error rates.
    
    \subsection{Method} \label{sec:DMP_method}
        
        Consider a stabiliser code encoding $k$ logical qubits into $n$ physical qubits, with code space spanned by the logical basis $\mathcal{B}_L=\{\ket{i_L}\}_{i=0}^{2^k-1}$.
        Let $X_i^L$, $i= 0, \dots, 2^k - 1$, be the logical bit flip operators such that $\ket{i_L} = X_i^L \ket{0_L}$.
        After multiple rounds of noisy stabiliser measurements, the state of the data qubits can be written as a classical mixture
        \begin{equation}
            \varrho = \sum_{s \in \{0, 1\}^{n_s}} p_s \varrho_s,
        \end{equation}
        where $p_s$ is the probability of observing syndrome $s$, $\varrho_s$ is the corresponding post-measurement state of the data qubits, and $n_s$ is the syndrome length.

        When a logical computational basis state is encoded initially, e.g., $\ket{0_L}$, the state $\varrho_s$ is diagonal in the logical basis up to a Pauli correction.
        We show in Appendix \ref{app:post-meas_state} that in that case, $\varrho_s$ can be expressed as 
        \begin{equation} \label{eq:rho_s_dec}
            \varrho_s = T \left(\sum_{i=0}^{2^k - 1} p_{s,i} X_i^L \ketbra{0_L} X_i^L \right) T, 
        \end{equation}
        where $T \in \mathcal{P}_n$ is determined by the syndrome and $p_{s,i}$ is a probability distribution over logical basis states.
        
        For a given syndrome $s$, the decoder identifies $X^L_i$ such that $\mathcal{R}_s = T X^L_i$ maps the most likely pure state back to the originally encoded state $\ket{0_L}$.
        This introduced the DMP decoding method.
        Denoting 
        \begin{equation}
        \label{eq:DMP_decoding_condition}
            w_{s} = \max_i p_{s, i},
        \end{equation}
        it corresponds to correcting the dominant component of $\varrho_s$.
        Since $T$ is unitary, the coefficients $p_{s,i}$ coincide with the eigenvalues of $\varrho_s$, and $w_s$ is its largest eigenvalue (or spectral norm).

                In the special case of a single logical qubit ($k=1$), the state $\varrho_s$ reduces to a mixture of two orthogonal logical states,
        \begin{equation}
            \varrho_s = p_{s,0} \mathcal{R}_s \ketbra{0_L} \mathcal{R}_s + p_{s, 1} \mathcal{R}_s \ketbra{1_L} \mathcal{R}_s,
        \end{equation}
        from which one finds that the largest eigenvalue can be expressed as $w_s = \nicefrac{1}{2}[1 + \sqrt{2 \tr(\varrho_s^2) - 1} ]$.
        This provides a convenient way to compute $w_s$ directly from the purity of $\varrho_s$.

    As an example, consider the distance-three repetition code, with codewords $\ket{0_L} = \ket{000}$ and $\ket{1_L} = \ket{111}$, subject to code-capacity noise, where each data qubit undergoes an $X$-error independently with probability $p$.
    Assuming the measured syndrome is $s=(1,0)$ after preparing the state $\ket{0_L}$, \cref{eq:rho_s_dec} reduces to
    \begin{subequations}
    \begin{align}
        \varrho_{10}
        & = 
        (1-p) \ketbra{100}{100}
        + p\ketbra{011}{011},
        \\
         &=
        X_1 \Bigl[
            (1-p) \ketbra{0_L}{0_L}
            + p X_L \ketbra{0_L}{0_L} X_L
        \Bigr] X_1,
        \label{eq:ex_rep_code_1}
    \end{align}
    \end{subequations}
    where $X_L = X_1 X_2 X_3$ is the single-logical-qubit $X$ operator, which maps $\ket{0_L}$ to $\ket{1_L}$, and where we fixed $T= X_1$ as the representative of $s = (1, 0)$.
    
    At small $p$, the DMP decoder identifies the leftmost term as dominant according to \cref{eq:DMP_decoding_condition}, since it has the highest probability.
    Consequently, it chooses $\mathcal{R}_{10} = X_1 {\id}_L$ as a recovery operation, correctly restoring the initial state $\ket{0_L}$ with probability $1-p$.

    \subsection{Estimation of the logical error rate}     \label{sec:DMP_estimation_log_err_rate}
        
        The conditional probability of a logical failure given syndrome $s$ is therefore 
        \begin{equation}
            p(\mathrm{fail} \mid s) = 1 - w_s,
        \end{equation}
        as all other components correspond to incorrect logical states.
        Averaging over all syndromes yields the logical error rate
        \begin{equation} \label{eq:ler_DMP}
            \eps_L = \sum_{s \in \{0, 1\}^{n_s}} p_s \, p(\mathrm{fail}\mid s).
        \end{equation}

    Evaluating \cref{eq:ler_DMP} exactly requires summing over all possible syndrome histories, whose number grows exponentially with the number of measurement rounds.
    In our implementation, this sum is organised as a depth-first search of a \emph{syndrome tree}, which provides a natural representation of all possible measurement outcomes.

    The root of the tree corresponds to the initial encoded state.
    Each level of the tree represents one round of stabiliser measurements (syndrome extraction), and each edge corresponds to a particular outcome of that round.
    A node at depth $t$ is therefore labelled by a \emph{partial syndrome history} $q \in \{0,1\}^{t(n-k)}$, and each leaf at depth $r+1$ corresponds to a full syndrome history $s \in \{0,1\}^{n_s}$.
    
    To each node $q$, we associate an \emph{unnormalised conditional state} $\varrho_q$, obtained by propagating the initial state through the noisy circuit and conditioning on the partial syndrome history $q$.
    A key property, which is directly exploited in the implementation, is that
    \begin{equation}
        p_q = \tr(\varrho_q)
    \end{equation}
    is exactly the total probability of all full syndrome histories descending from that node.

    The logical error rate is then obtained by traversing the tree in a depth-first manner.
    For each leaf $s$, we compute the contribution $p_s \, p(\mathrm{fail}\mid s)$ and sum over all leaves.
    For classical codes and depolarising noise models, the matrices $\varrho_s$ are already diagonal, which simplifies the evaluation of $p(\mathrm{fail} \mid s)$ (see \cref{sec:DMP_method}).

    In the repetition-code example of \cref{sec:DMP_method}, we have $w_{10} = 1-p$ for the syndrome $s=(1,0)$, yielding
    \begin{equation}
        p_{10} \, p(\mathrm{fail}\mid 10) = p^2(1-p).
    \end{equation}

    We also emphasise that all syndrome histories are explicitly and exhaustively explored in this procedure. 
    As a result, the logical error rate in \cref{eq:ler_DMP} is free of statistical uncertainty and not restricted to any particular range of physical error rates.

    \subsection{Truncated tree exploration}     \label{sec:DMP_truncated_tree_exploration}

    The size of the syndrome tree scales exponentially with the number of rounds, making exhaustive traversal quickly intractable.
    To reach regimes with a larger number of rounds, we exploit the fact that many syndrome branches are unlikely ($p_q = \tr(\varrho_q)$ is small) and therefore contribute little to the logical error rate.
    Specifically, at each node $q$, we compare $p_q$ to a predefined cutoff.
    If $p_q$ falls below this threshold, the entire subtree rooted at $q$ is pruned, and no further exploration is performed along that branch. 
    
    This is motivated by the fact that for the repetition code, the fraction of syndrome histories that significantly contributes to the logical error rate decreases rapidly with the syndrome length.
    For $d=3$ and $r = 3$ at physical error rate $p=10^{-3}$, keeping $\sim 25\%$ of syndrome histories suffices to achieve $1\%$ accuracy on $\eps_L$.
    For the distance-5 repetition code at the same noise level, this fraction drops to $\sim 5 \%$ for $r=3$, and to $0.1\%$ for $r = 5$ (see Appendix \ref{app:repetition_code_syndrome_expl}).
    This suggests that an exhaustive sum is highly inefficient, as only a small subset of syndrome histories contributes significantly to the logical error rate.
    The challenge, however, is that these relevant syndrome histories are not known a priori and must be identified during the exploration itself.

    Pruning introduces a controlled approximation, but it also naturally provides rigorous bounds on the logical error rate.

    Let $\Omega$ denote the set of fully explored leaves (i.e., complete syndrome histories), and $\Pi$ the set of pruned nodes.
    Each term of the sum in \Cref{eq:ler_DMP} being positive, the contribution from the explored branches defines a lower bound
    \begin{equation}
        \eps_L^{\mathrm{low}} 
        = 
        \sum_{s \in \Omega} p_s \, p(\mathrm{fail}\mid s).
    \end{equation}
    The missing contribution from pruned subtrees can be upper bounded using the fact that, for any syndrome, the failure probability satisfies $p(\mathrm{fail}\mid s) \leq 1 - \nicefrac{1}{2^k}$, since at least one logical state is correctable.
    This yields the upper bound
    \begin{equation}
        \eps_L^{\mathrm{up}}
        =
        \eps_L^{\mathrm{low}}
        + \left( 1 - \frac{1}{2^k} \right)
        \sum_{q \in \Pi} p_q,
    \end{equation}
    Altogether,
    \begin{equation}
        \eps_L^{\mathrm{low}}
        \leq
        \eps_L
        \leq
        \eps_L^{\mathrm{up}}.
    \end{equation}
    In practice, the lower bound typically converges much faster than the upper bound. 
    This is because the dominant contributions to the logical error rate arise from a relatively small subset of high-probability syndrome histories, whereas the upper bound must account for all unexplored branches without having information on their failure probability, which is typically much lower than $1 - \nicefrac{1}{2^k}$.
    Further statistical properties of the syndrome distribution are discussed in \cref{sec:syndrome_statistics}.

\section{Decoder performance comparison}

    With the DMP decoding method now introduced, we proceed to benchmark it against state-of-the-art decoding algorithms.
    All reference decoders are evaluated using the standard Monte Carlo sampling approach to estimate logical error rates, in contrast to DMP, which does not require sampling.

    \subsection{Error sampling} \label{sec:error_sampling}
    For the estimation of the logical error rate through error sampling, one simulates repeated executions of the noisy circuit.
    For each run, errors are sampled according to the chosen noise model and applied throughout state preparation, gate operations, and measurements.
    
    For circuits composed of Clifford operations, such simulations can be performed efficiently using stabiliser-based techniques~\cite{gottesman1998heisenberg, aaronson2004improved, anders2006fast}.
    In practice, highly optimised implementations such as \texttt{Stim}~\cite{gidney2021stim} are commonly used.
    
    After simulating $N$ noisy trajectories, of which $N_e$ lead to a logical error, the logical error rate is estimated as 
    \begin{equation}
        \widehat{\eps}_L = \frac{N_e}{N}.
    \end{equation}
    This estimator is unbiased and approximates the exact logical error rate defined in \cref{sec:DMP_estimation_log_err_rate} (see Appendix \ref{app:log_err_equiv} for the equivalence between the two formulations).
    
    Sampling-based approaches scale to much larger systems but carry statistical uncertainty, as opposed to the density matrix approach.
    The controlled values obtained in \cref{sec:DMP_estimation_log_err_rate} therefore serve as a benchmark for assessing the performance of practical decoding strategies.

    \subsection{Benchmarking decoder accuracy} \label{sec:benchmark_accuracy}

    Using the logical error rates obtained with the DMP decoder, we can directly quantify the suboptimality of practical decoding algorithms.
    In this section, we compare several widely used decoders, namely MWPM~\cite{higgott2025sparse}, BP+OSD~\cite{roffe2020decoding}, Planar~\cite{feng2025planar}, and the Tesseract decoder~\cite{beni2025tesseract}.
    
    All simulations are performed using a circuit-level noise model parametrised by a physical error rate $p$, which sets the probability of each elementary error occurrence.
    In this model, qubit initialisation and resets are faulty, Clifford gates are followed by stochastic Pauli errors, unused qubits are subject to idle errors, and measurement outcomes are noisy.
    
    For classical codes, this reduces to bit-flip noise at all circuit locations: reset errors, idling errors, and measurement errors occur with probability $p$, while each $\textsc{cnot}$ is followed by one of $\{X_i, X_j, X_iX_j\}$ with probability $\nicefrac{p}{3}$. 

    We benchmark these decoders on two families of stabiliser codes.
    First, we consider the repetition code, which serves as a reference case.
    Its decoding properties are well understood, and near-optimal performance is achieved by standard decoders~\cite{cao2025exact}.
    As expected, we find only minor differences between heuristic and ML decoding in this setting (see Appendix \ref{app:repetition_code_log_error_rate}).
    
\begin{figure}
        \centering
        \includegraphics[width=0.75\linewidth]{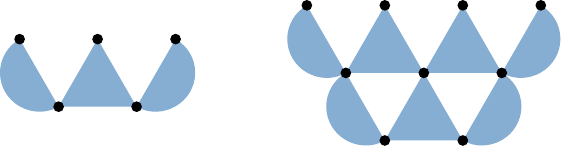}
        \caption{Stabilisers of the triangular code for distance three (left) and five (right), corresponding to $\mathcal{T}_1$ and $\mathcal{T}_2$, respectively.
        The blue plaquettes represent $Z$-stabilisers acting on the data qubits at their vertices.}
        \label{fig:cell_auto_stabilisers}
    \end{figure}

    Our main case study is instead a family of cellular automaton codes introduced in Refs.~\cite{ruiz2025private, ruiz2025ldpc}.
    We refer to this family as the triangular code, since the stabilisers are defined recursively on a triangular lattice as illustrated in \cref{fig:cell_auto_stabilisers}.
    
    The triangular code $\mathcal{T}$ is a family of classical low-density parity check (LDPC) codes that encodes two logical qubits and is indexed by an integer $\ell \geq 1$.
    The code parameters of $\mathcal{T}_\ell$ are $[n(\ell), 2, d(\ell)]$, where
    \begin{equation}
        n(\ell) = \frac{(\ell + 2)(\ell + 3)}{2} - 1.
    \end{equation}
    The distance $d(\ell)$ is given by the number of odd entries in the corresponding row of Pascal’s triangle, known as Gould’s sequence~\cite{oeisA001316}.
    The odd entries of Pascal's triangle form a fractal, known as Sierpiński triangle \cite{wolfram1984geometry}.
    As a consequence, the logical operators have a fractal support and form a Sierpiński triangle (see Appendix \ref{app:cell_auto} for the details).
    In contrast to the repetition code \cite{cao2025exact}, efficient near-optimal decoding strategies for this family are not well understood, making it a natural benchmark for assessing decoder suboptimality.

\begin{figure*}
        \centering
        \includegraphics[width=0.9\linewidth]{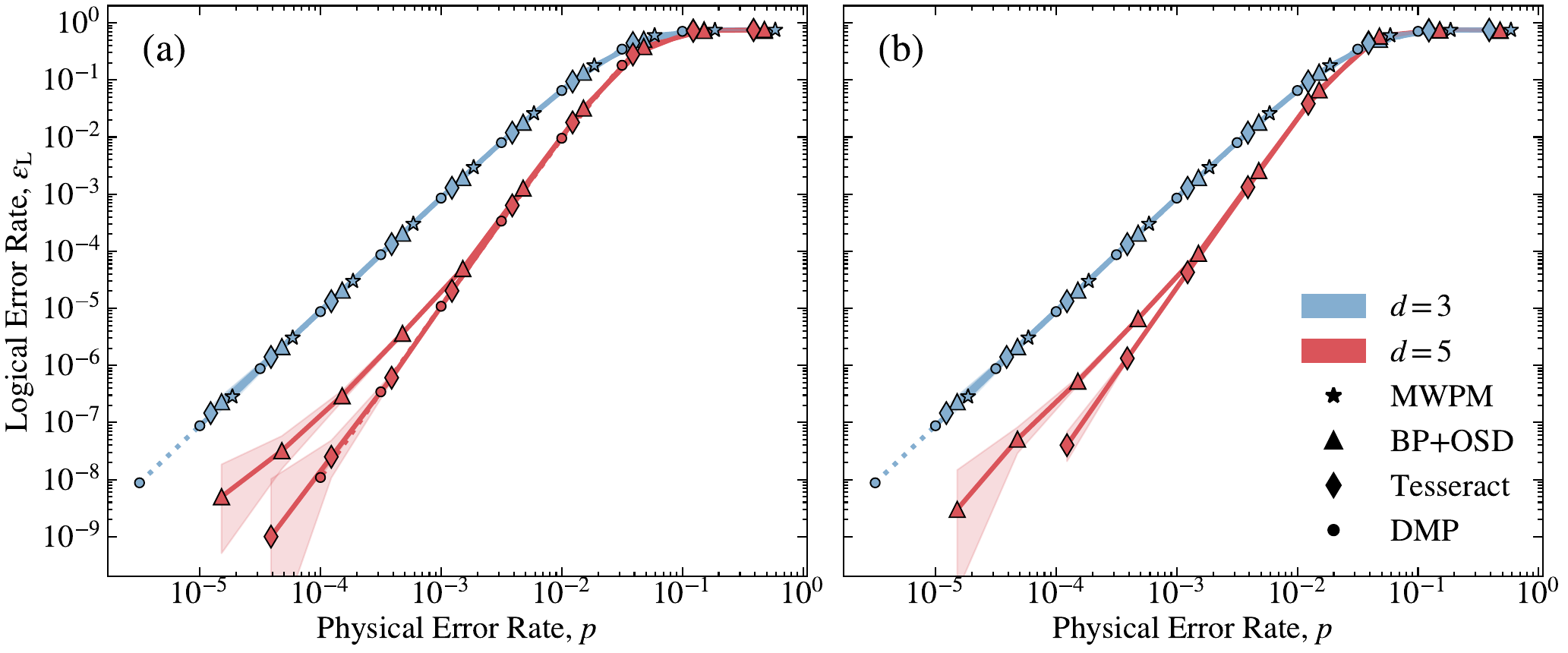}
        \caption{%
            Logical error rate as a function of the physical error rate $p$ for the triangular code $\mathcal{T}$ with (a) $r=3$ and (b) $r=d$ rounds, using various decoders at distances $d=3$ and $d=5$. 
            Shaded regions indicate statistical error bars obtained from $N=10^9$ samples.
            The DMP simulations were performed without pruning for $d=3$. For $d=5$, pruning was used and the plotted values correspond to the lower bound $\eps_L^{\mathrm{low}}$. 
            The cutoff was chosen such that the gap $\abs{\eps_L^{\mathrm{up}}-\eps_L^{\mathrm{low}}}$ remained around $10\%$ of $\eps_L^{\mathrm{low}}$, making the gap negligible on the scale of the figure.
            The computation of all the DMP data points took over $125 \, 000$ hours on a single work station CPU.
            For $d=5$ and $r=d$, the total number of syndrome histories is $2^{42} \approx 4.4 \times 10^{12}$, making DMP simulations computationally prohibitive even with pruning (see Appendix \ref{app:repetition_code_syndrome_expl} for a discussion).
        }
        \label{fig:results_combined_ldpc}
    \end{figure*}

    Figure \ref{fig:results_combined_ldpc} shows the logical error rate of the triangular code for distances $d=3$ and $d=5$, using different decoders and varying the number of syndrome-extraction rounds.
    The results reveal substantially larger differences between decoders than in the repetition code case.
    In particular, BP+OSD deviates noticeably from the other decoders at low physical error rates, suggesting that belief propagation struggles to capture the relevant correlations in this regime.    
    Interestingly, this discrepancy appears at $d=5$ but not at $d=3$.
    We attribute this to the increased decoding complexity in the former case. 
    As illustrated in Appendix \ref{app:cell_auto}, the matching graph---and consequently the associated Tanner graph---of the distance-5 triangular code contains hyperedges, that is, edges incident on more than two vertices. 
    This is in contrast to the distance-3 case, where each edge is incident to at most two vertices.
    This prevents the use of some efficient decoders like MWPM, and may affect BP+OSD performance.
    A detailed investigation of this effect is left to future work.
    At larger physical error rates, however, all decoders again behave nearly optimally.
    
    Among the considered decoders, Tesseract achieves the best agreement with ML decoding at low physical error rates, but its current implementation becomes computationally expensive as $p$ increases.
    BP+OSD instead provides a favourable compromise between runtime and decoding performance over a broader range of parameters.

    \subsection{Syndrome concentration and pruning efficiency} \label{sec:syndrome_statistics}
        
    The main limitation of the DMP decoder is the exponential growth of the syndrome space.
    Indeed, the total number of syndrome histories scales as $2^{n_s}$, where $n_s$ is the syndrome length.
    The pruning strategy introduced in \cref{sec:DMP_truncated_tree_exploration} is therefore essential to access larger system sizes.

\begin{figure}
        \centering
        \includegraphics[width=0.95\linewidth]{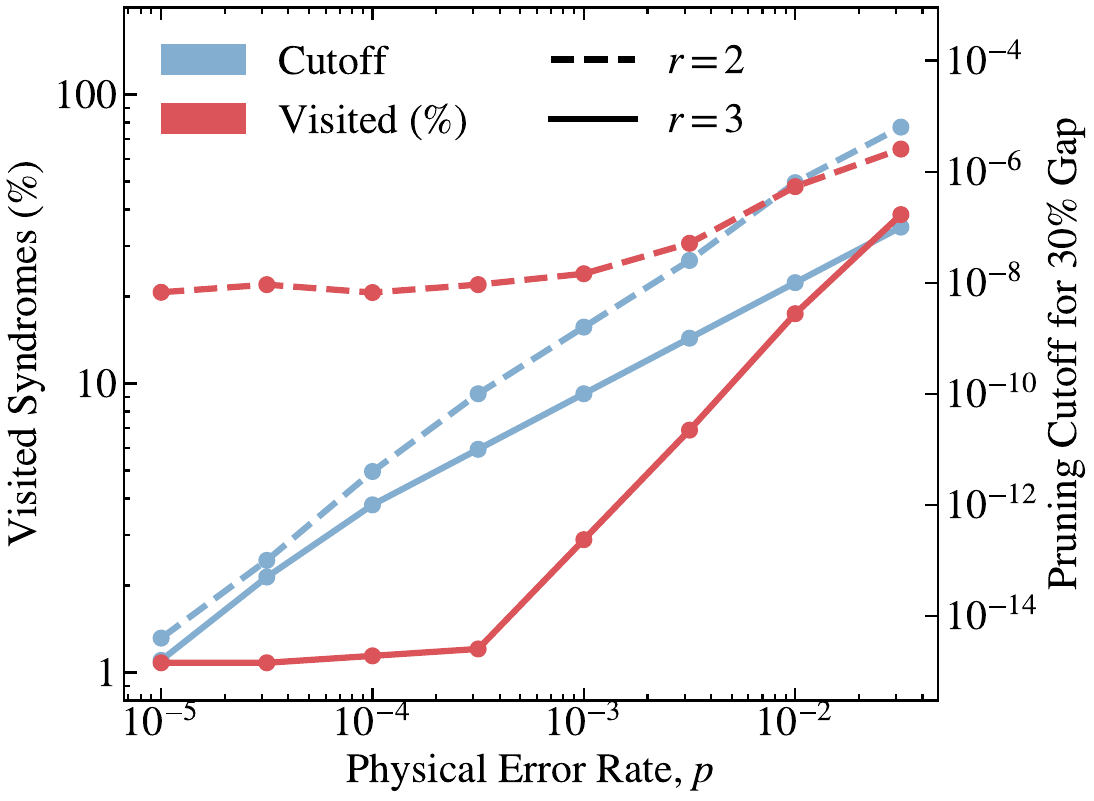}
        \caption{%
        Fraction of syndrome histories explored to obtain a $30\%$ gap between the upper and lower bounds on the logical error rate (red, left axis) for the distance-five triangular code $\mathcal{T}_2$
        with $r=2$ and $r=3$ rounds.
        The corresponding pruning cutoff is shown on the right axis (blue).
        For a number of rounds $r=2$ and $r=3$, the total number of syndrome histories is $2^{21} > 2\times10^6$ and $2^{28} > 2\times10^8$, respectively.
        }
        \label{fig:percent_visited}
    \end{figure}

    A key observation is that, at low physical error rates, only a small fraction of syndrome histories contributes significantly to the logical error rate.
    To quantify this effect, we compute the fraction of syndrome histories that must be explored in order to achieve $\abs{\eps_L^{\mathrm{up}} - \eps_L^{\mathrm{low}}} \approx 0.3 \ \eps_L^{\mathrm{low}}$ for the distance-five triangular code $\mathcal{T}_2$.
    We stress that this $30\%$ gap is used only to study syndrome concentration, not for logical error rate estimation.
    Figure \ref{fig:percent_visited} shows the resulting fraction of explored syndrome histories as a function of the physical error rate.
    We observe that the fraction of relevant syndrome histories rapidly decreases and appears to saturate as the physical error rate decreases.
    For $r=2$, the saturation occurs around $20\%$, while for $r=3$ it drops to approximately $1\%$.
    
    In practice, the computational cost remains substantial despite the reduction in the number of explored syndrome histories.
    For the distance-five triangular code with $r=3$, the computation of the extremal points shown in \cref{fig:percent_visited} at $p=10^{-1.5}$ and $p=10^{-5}$ respectively required around $30 \, 000$ and $1400$ hours on a single workstation CPU.
    In this case, the density matrix of the data and ancillary qubits has dimension $2^{16}=65 \, 536$.
    Since the post-measurement state is diagonal in the computational basis, it is stored in sparse format, such that memory consumption remains moderate.
    The current limitation is therefore computational time rather than memory usage.
    We also note that the exploration of independent branches of the syndrome tree can be parallelised straightforwardly across multiple CPUs.

    This behaviour is consistent with the intuition that, at low physical error rates, the logical error rate is dominated by a relatively small number of low-weight error histories.
    At larger physical error rates, contributions become more broadly distributed across syndrome space, making pruning less effective.

    These observations suggest the existence of a crossover physical error rate below which the fraction of relevant syndrome histories becomes approximately constant. 
    While estimating this crossover analytically remains an open problem, it may be related to the waterfall regime discussed in Ref.~\cite{gu2026scalable}.

\section{Conclusion and outlook}
    
    In this work, we presented a density matrix framework for ML decoding under circuit-level noise, allowing us to compute logical error rates together with rigorous upper and lower bounds.
    We used this framework to benchmark heuristic decoders and investigate the fraction of relevant syndrome histories contributing to the logical error rate.

    Whereas state-of-the-art decoders are nearly optimal for the repetition code \cite{cao2025exact}, larger deviations from ML decoding are observed for the triangular code.
    More broadly, our results show that decoder performance can differ substantially from optimal decoding even at small distances when realistic circuit-level noise and time-like degeneracies are taken into account.
    Moreover, the pruning method developed here shows that, at low physical error rates, only a small fraction of syndrome histories contributes significantly to the logical error rate.
    This suggests that future decoding strategies could focus on the relevant regions of syndrome space for a given physical error rate instead of treating all syndromes equally.

    Despite the pruning strategy, the computational cost of exact decoding still grows exponentially with the number of syndrome-extraction rounds, such that the method remains limited to relatively small codes and distances.
    The examples considered here were restricted to effectively classical codes, although the framework itself works for general quantum codes and more general noise models.

    Several directions could improve the scalability of the approach.
    In particular, extending the benchmarks to larger distances and numbers of rounds could be possible through sampling over syndrome histories instead of the exhaustive sum of \cref{eq:ler_DMP}.
    This could be combined with a matrix-product-state (MPS)~\cite{orus2014practical}, such that the MPS acts as a decoder when given a syndrome history \cite{cussenot2025quantum}.
    More generally, DMP decoding could provide useful insight into syndrome statistics and be used to develop more efficient syndrome-based sampling methods.

    Beyond decoder benchmarking, exact ML decoding under complex noise models may also be useful for the off-line computation of logical error rates in proof-of-principle quantum error-correction experiments.

\section*{Acknowledgements}
    We thank G.~Fregona, C.~de~Gois, E.~Gouzien, B.~Grivet, J.~Guillaud, J.~Houdayer, H.~Jacinto, D.~Ruiz, and X.~Valcarce for useful discussions.
    This work was supported by the French National program Programme d’Investissement d’Avenir, IRT Nanoelec, reference ANR-10-AIRT-05 and by the Agence Nationale de la Recherche in the framework of France 2030, reference ANR-22-PETQ-0007.
    P. C. acknowledges funding from the French Direction Générale de l’Armement (DGA).

\bibliographystyle{apsrev4-2}
\bibliography{bib_clean}

\newpage

\onecolumngrid

\appendix

\section{Density matrix formalism} \label{app:post-meas_state}
    In this appendix, we show that the conditional post-measurement states $\varrho_s$ after the encoding of $\ket{0_L}$ can be written as
    \begin{equation} 
        \varrho_s = T \left(\sum_{i=0}^{2^k - 1} p_{s,i} X_i^L \ketbra{0_L} X_i^L \right) T, 
    \end{equation}
    thus recovering \cref{eq:rho_s_dec} of the main text.
    
    Let $\mathcal{E}$ denote the set of all $n$-Pauli errors that can affect the logical qubit system. 
    Just before the final stabiliser measurements, the state of the system can be written as a classical mixture over all possible error realisations,
    \begin{equation}
        \varrho = \sum_{E \in \mathcal{E}} p_E \, E \ketbra{\Psi_L} E^\dagger,
    \end{equation}
    where $\ket{\Psi_L} \in \mathcal{C}$ is the initially encoded logical state.
    
    Measuring the $n-k$ generators yields a syndrome $s = (s_1, \dots, s_{n-k})$ and projects the state onto the corresponding joint eigenspace. 
    Denote $\Pi_s$ the projector onto this subspace, the post-measurement state is  
    \begin{equation}
        \varrho \mapsto \varrho_s = \frac{\Pi_s \varrho \Pi_s}{p_s},
    \end{equation}
    where $p_s = \tr(\Pi_s \varrho)$ is the probability of observing the syndrome $s$.

    To characterise which errors contribute to $\varrho_s$, it is useful to examine how Pauli errors interact with the stabiliser projectors.
    For any Pauli error $E$, one has
    \begin{equation}
        \Pi_s E = E \, \Pi_{s^\prime},
    \end{equation}
    where $s'$ is the syndrome obtained by conjugating the stabilisers with $E$.
    More explicitly, for each generator $g_i$, one has
    \begin{equation}
        g_i E = x_i E g_i, \qquad x_i \in {\pm 1},
    \end{equation}
    so that the syndrome transforms as $s'_i = s_i x_i$.
    We say that an error $E$ is \emph{compatible} with the syndrome $s$ if it produces that syndrome when acting on a code state, i.e., if $s_i = x_i$ for all $i \in \{1, \dots, n-k\}$.
    In that case, 
    \begin{equation}
        \Pi_s E \ket{\Psi_L} = E \Pi_{s^\prime} \ket{\Psi_L} = E \ket{\Psi_L}.
    \end{equation}
    If $E$ is not compatible with $s$, then $\Pi_s E \ket{\Psi_L} = E \Pi_{s^\prime} \ket{\Psi_L} = 0$.
    
    From this observation, it follows that only compatible errors contribute to the post-measurement state, yielding
    \begin{equation}
        \Pi_s \varrho \Pi_s = \sum_{E \in \mathcal{E}_s} p_{E} E \ketbra{\Psi_L} E^\dagger,
    \end{equation}
    where $\mathcal{E}_s$ is the subset of $\mathcal{E}$ consisting of errors compatible with $s$.
    Normalising, we obtain
    \begin{equation} \label{eq:rho_s}
        \varrho_s = \sum_{E \in \mathcal{E}_s} p(E \mid s) E \ketbra{\Psi_L} E^\dagger,
    \end{equation}
    with $p(E \mid s) = \nicefrac{p_E}{p_s}$.
    Thus, $\varrho_s$ is the conditional mixture of all the error trajectories compatible with the syndrome $s$.
    
    Next, we use the TLS decomposition introduced in the main text. 
    Up to a phase, any Pauli error can be written as
    \begin{equation}
        E = \, T \, L \, S,
    \end{equation}
    where $S \in \mathcal{S}$ is a stabiliser, $L$ is a logical operator, and $T$ is a pure error determined by the syndrome.
    
    For a fixed syndrome $s$, the pure error $T$ is fixed (once a convention is chosen), so all compatible errors differ only by their logical and stabiliser components. 
    Substituting into \cref{eq:rho_s}, we obtain
    \begin{equation}
        \varrho_s = T \left(\sum_{L, S} p(E \mid s) \, L S \ketbra{\Psi_L} S^\dagger L^\dagger \right) T^\dagger.
    \end{equation}
    where the sum runs over all $L$ and $S$ such that $TLS \in \mathcal{E}_s$.

    Since stabilisers act trivially on the code space, $S \ket{\Psi_L} = \ket{\Psi_L}$, this simplifies to
    \begin{equation}
        \varrho_s = T \left( \sum_{L} p_L \, L \ketbra{\Psi_L} L^\dagger \right) T^\dagger,
    \end{equation}
    where $p_L$ is the total probability of all errors with logical component $L$.

    The role of decoding is precisely to identify the most likely logical operator. 
    Let
    \begin{equation}
        \label{eq:DMP_ML_decoding_argmax_pL}
        L_M = \argmax_{L} p_L,
    \end{equation}
    and define the recovery operation
    \begin{equation}
        \mathcal{R} = T L_M.
    \end{equation}
    Applying this recovery, we obtain
    \begin{equation}
        \varrho_s = \mathcal{R} \left( \sum_{L} p_L L_M L \ketbra{\Psi_L} L^\dagger L_M^\dagger \right) \mathcal{R}^\dagger.
    \end{equation}

    Finally, consider the case where the encoded state is a logical computational basis state, e.g., $\ket{\Psi_L} = \ket{0_L}$.
    Logical operators then either act trivially or map $\ket{0_L}$ to another basis state $\ket{i_L}$ within the logical computational basis $\mathcal{B}_L$.
    As a result, the post-measurement state takes the form
    \begin{equation}
        \varrho_s = \mathcal{R} \left(\sum_{i=0}^{2^k - 1} p_i \ketbra{i_L} \right) \mathcal{R}^\dagger.
    \end{equation}
    We thus conclude that, when the encoded state is a computational basis state, the conditional state $\varrho_s$ is—up to the recovery operation—diagonal in the logical computational basis.

\section{Equivalence of sampling and syndrome-based formulations} \label{app:log_err_equiv}
    In this appendix, we show that the sampling-based and density matrix approaches compute the same logical error rate.
    
    Let $m$ denote a full \emph{error pattern}, i.e., a realisation of stochastic errors occurring throughout the circuit, with probability $p(m)$.
    Given a decoder, define
    \begin{equation}
        f_{\textrm{dec}}(m) = 
        \begin{cases}
            1 \quad \text{if $m$ leads to a logical error}, \\
            0 \quad \text{otherwise.}
        \end{cases}
    \end{equation}
    The logical error probability reads
    \begin{equation} \label{eq:app:log_err_patterns}
        \eps_L = \sum_{m} p(m) \, f_{\textrm{dec}}(m)
    \end{equation}
    where the sum runs over all error patterns and is the quantity estimated in Monte Carlo simulations.
    
    On the other hand, the density matrix approach groups error patterns according to their syndrome.
    Denoting by $s$ a syndrome history and by ${m_s}$ all error patterns leading to $s$, one has
    \begin{equation}
        p(\mathrm{fail} \mid s) = \sum_{m_s} f_{\mathrm{dec}}(m_s) \, p(m_s \mid s) 
    \end{equation}
    where the sum runs over all error patterns leading to $s$.
    Using $p(m_s \mid s) = \nicefrac{p(m_s)}{p(s)}$, the logical error rate becomes
    \begin{equation}
        \eps_L = \sum_{s} p(s) \, p(\mathrm{fail} \mid s) = \sum_{s} \sum_{m_s} p(m_s) \, f_{\mathrm{dec}}(m_s),
    \end{equation}
    which is identical to \cref{eq:app:log_err_patterns}.
    
    This shows that both approaches compute the same quantity: the density matrix method performs an exact summation over all error patterns grouped by syndrome, while sampling-based methods approximate the same sum via Monte Carlo.

\section{Repetition code} \label{app:repetition_code}
    In this appendix, we review the repetition code that acts as a base for the exploration of the relevant syndrome histories (Appendix \ref{app:repetition_code_syndrome_expl}), as well as a sanity check for the logical error rate evaluations (Appendix \ref{app:repetition_code_log_error_rate}).

    \subsection{Syndrome exploration} \label{app:repetition_code_syndrome_expl}
    The main limitation of the DMP method lies in the exponential growth of the number of syndromes to be explored. 
    Indeed, the total number of syndromes is $2^{n_s}$, where $n_s = (r+1) (n-k)$ denotes the syndrome length.
    To extend the applicability of our method, we propose to restrict the exploration to relevant syndrome histories only. 
    In this appendix, we show that for the repetition code, a small subset of all syndromes is sufficient to accurately approximate the logical error rate.
    
\begin{figure}
        \centering
        \includegraphics[width=.9\linewidth]{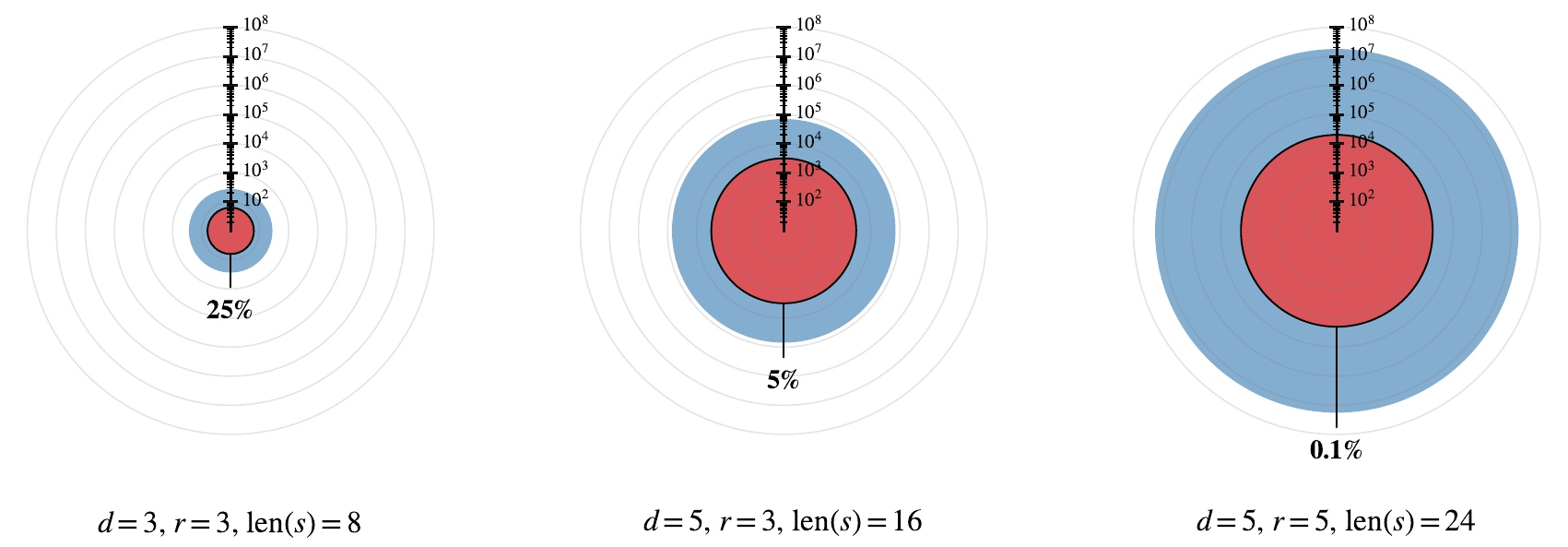}
        \caption{%
        Total number of syndrome histories (blue) and minimal number required to approximate the logical error rate within $1\%$ relative error (red) for the repetition code across various code parameters. 
        The relevant syndromes are selected based on their largest contributions to the logical error rate, $p_s \, p(\mathrm{fail}\mid s)$, and the percentage indicates their fraction relative to the total.
        The physical error rate is $p=10^{-3}$.}
        \label{fig_syndromes_explored_vs_all}
    \end{figure}

    To that end, we compare the total number of syndrome histories with the smallest number required to approximate the logical error rate within a $1\%$ relative error.
    The smallest subset is obtained by selecting the syndromes with the largest contributions to the logical error rate, $p_s \, p(\mathrm{fail}\mid s)$ (see \cref{eq:ler_DMP} of the main text).
    The required fraction is plotted in \cref{fig_syndromes_explored_vs_all} for various code distances $d$ and numbers of rounds $r$. 
    \begin{figure}
        \centering
        \includegraphics[width=.4\linewidth]{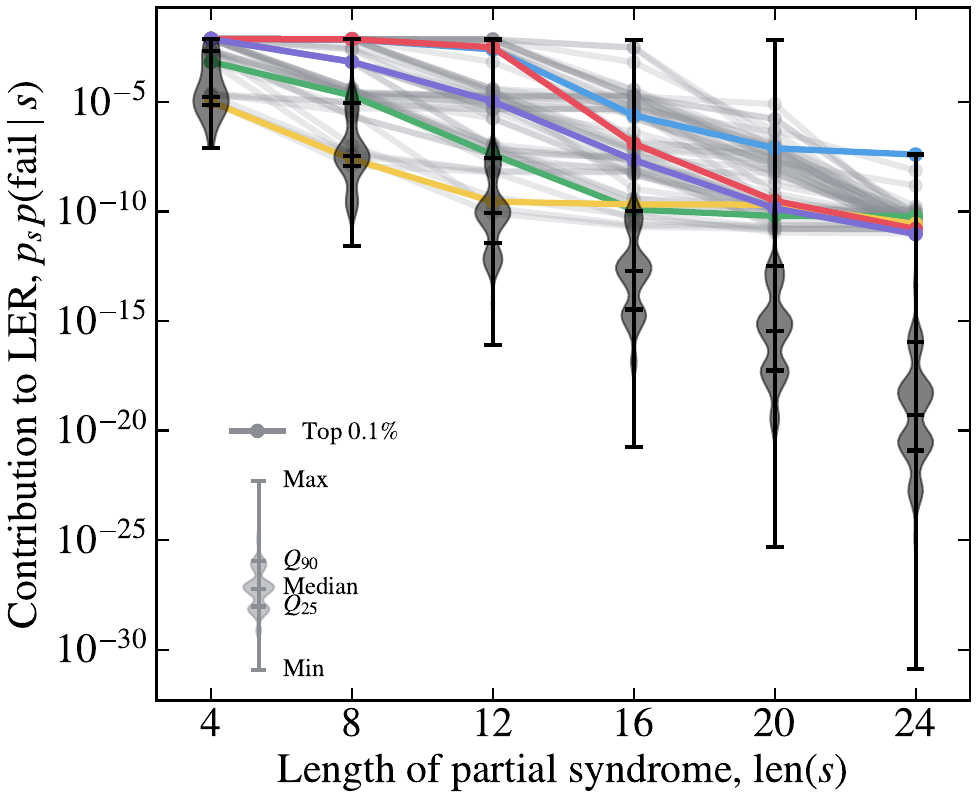}
        \caption{Representative paths in the syndrome tree for the distance-5 repetition code ($r=5$ rounds, which leads to a syndrome history of length $(r+1)(d-1)=24$).
        Black violin plots show the full distribution of syndromes. Grey paths represent typical paths from the top $0.1\%$ of syndromes contributing most to the logical error rate $\eps_L$, while coloured paths are selected examples from these grey paths.
        The yellow path illustrates that, after one and two rounds (syndrome lengths $4$ and $8$), most syndromes contribute more than it, yet it becomes one of the highest-contributing paths after all $d$ rounds.
        The physical error rate is $p=10^{-3}$.}
        \label{fig_violon_plot_syndromes}
    \end{figure}

    Although the number of relevant syndrome histories increases with system size, their fraction relative to the total decreases significantly. 
    For instance, for $d=5$ and $r=5$, selecting the $20\,000$ syndromes with the highest contribution--corresponding to only $0.1\%$ of all syndromes--is sufficient to recover the logical error rate within $1\%$ relative error.
    
    At first sight, these results suggest that the exploration could be drastically reduced by focusing only on a small subset of syndrome histories. 
    The key challenge, however, is that these highly contributing syndromes are not known in advance and must be identified during the exploration itself. 
    This observation motivates the introduction of a cutoff for syndrome exploration as described in \cref{sec:DMP_truncated_tree_exploration} of the main text.

    A natural question is whether the syndrome histories that contribute most to the logical error rate can already be identified after only a few syndrome-extraction rounds. 
    If this were the case, one could prune large parts of the syndrome tree at an early stage while still retaining the most relevant contributions.

    To investigate this, \cref{fig_violon_plot_syndromes} shows representative paths in the syndrome tree for the distance-$5$ repetition code with $r=5$. We consider typical paths among the top $0.1\%$ of syndrome histories contributing most to the logical error rate and compare them with the full distribution shown as black violin plots. 
    The grey lines represent typical high-contributing paths, while a few examples are highlighted in colour.
    
    The figure shows that the paths contributing the most after the final round are not necessarily among the most important ones at intermediate stages of the exploration. 
    For example, after one and two rounds (corresponding to syndrome lengths $4$ and $8$), more than half of all syndromes contribute more to the logical error rate than the yellow path. 
    Nevertheless, after all rounds have been completed, this same path belongs to the top $0.1\%$ of syndrome histories contributing the most to the logical error rate.
    Conversely, the blue and red highlighted paths remain in the most significant fraction across all rounds.
    
    This illustrates that the contribution of a syndrome history is difficult to predict from partial information alone. 
    A path that appears unimportant after a few rounds can later become one of the dominant contributions. 
    This likely explains why the pruning method provides only a limited reduction of the computational cost despite the fact that only a small fraction of syndrome histories ultimately contributes significantly to the logical error rate.

    \subsection{Logical error rate} \label{app:repetition_code_log_error_rate}
    
\begin{figure}
        \centering
            \centering
            \includegraphics[width=.4\linewidth]{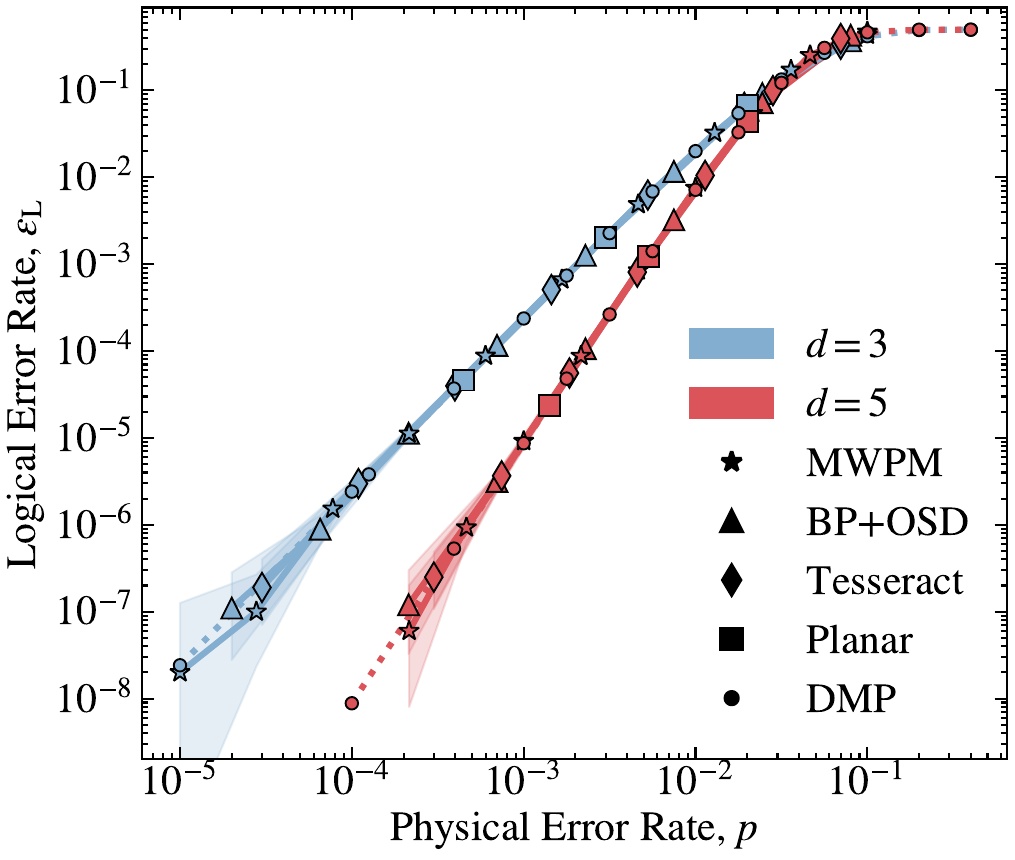}
        \caption{
            Logical error rate as a function of the physical error rate $p$ for the repetition code with $r=d$ rounds, using various decoders at distances $d=3$ and $d=5$. 
            Shaded regions indicate statistical error bars.
            We used $N=10^8$ shots except for the \texttt{Planar} decoder for which we used $N=10^7$ shots. 
            The \texttt{Planar} and DMP decoders are ML, in contrast to the others. 
            The DMP simulations were realised without pruning and are therefore exact to machine precision.
        }
            \label{fig:results_repetition_large}
    \end{figure}

    For the repetition code with $r=d$ rounds of syndrome extraction, all decoders yield very similar logical error rates across the full range of physical error rates (see \cref{fig:results_repetition_large}).
    This is expected, as the repetition code has a simpler matching graph, for which MWPM is known to provide near-optimal decoding.
    
    Since all decoders provide comparable accuracy, the most practical choice is MWPM: it matches the DMP decoder well while computing complete performance curves within minutes on a laptop, unlike BP+OSD and Tesseract, whose runtimes are an order of magnitude slower.
    Although the Planar decoder has low theoretical complexity, its Python implementation makes it so far one of the slowest methods in practice for the code sizes we implemented.

\section{Triangular code details} \label{app:cell_auto}

    \begin{figure}
        \centering
        \includegraphics[width=0.65\linewidth]{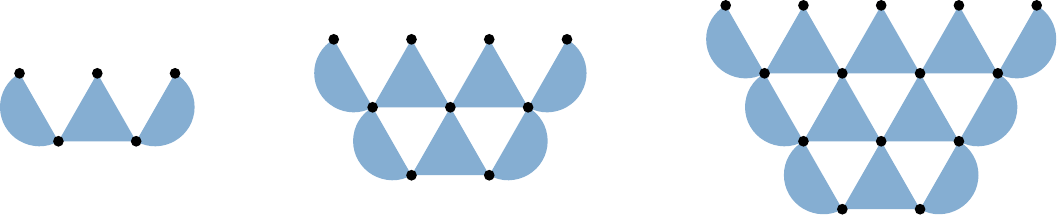}
        \caption{First three elements of the triangular code family $\mathcal{T}_1$, $\mathcal{T}_2$ and $\mathcal{T}_3$.
        The blue plaquettes represent $Z$-stabilisers acting on the data qubits at their vertices.}
        \label{fig:app:stab_cell_auto}
    \end{figure}

In this appendix, we describe the triangular code studied in this work.
We review the code parameters of the family and their logical operators, present the syndrome-extraction circuits for the distance-three and distance-five cases, and show the associated matching graphs under circuit-level noise.

The triangular code is a classical code, meaning that it corrects only one type of error---either bit-flip or phase-flip, depending on the chosen basis. 
Its stabilisers have either three-body support (triangles) or two-body support (half-disks) and are arranged in a triangular geometry. 
The iterative construction of the first three members of the code family is shown in \cref{fig:app:stab_cell_auto}, where each plaquette represents a $Z$-stabiliser acting on the data qubits at its vertices.

Each successive row contains one more data qubit than the row below.
The code family is constructed iteratively by adding a new row of data qubits at the top, which introduces a new layer of stabilisers: two-body stabilisers at the boundaries and three-body stabilisers in the bulk.

    \subsection{Code parameters}

    The triangular code family $\mathcal{T}$ is parametrised by an integer $\ell \geq 1$. 
    We denote its members by $\mathcal{T}_\ell$, where $\ell$ specifies the number of stabiliser layers.
    With this convention, \cref{fig:app:stab_cell_auto} shows $\mathcal{T}_1$, $\mathcal{T}_2$ and $\mathcal{T}_3$.
    The code parameters $[n(\ell), k, d(\ell)]$ of an arbitrary triangular code $\mathcal{T}_\ell$ are defined by
    \begin{align}
        n(\ell) &= \frac{(\ell + 2)(\ell + 3)}{2} - 1, \\
        k&=2,\\
        d(\ell) &= G_{\ell+1},
    \end{align}
    where $\{G_\ell\}_{\ell \in \mathbb{N}}$ denotes the partial sums of Gould's sequence~\cite{oeisA001316},
    \begin{equation}
        0, 1, 3, 5, 9, 11, 15, 19, 27, 29, 33, 37, 45, \dots
    \end{equation}
    which counts all odd numbers in the first $\ell+2$ rows of Pascal's triangle.
    The values scale as $d(\ell) \sim \ell^{\log_2(3)}$, so, the classical BPT bound~\cite{bravyi2010tradeoffs} is not saturated.
    Indeed,
    \begin{equation}
        k\sqrt{d} \sim 2 \, \ell^{0.7925} \ll \mathcal{O}(n) = \ell^2.
    \end{equation}

    In the next section, we examine the connection between the code distance of this family and Gould's sequence through an analysis of its logical operators.

    \subsection{Logical operators and Gould's sequence}
    
   To compute the distance of the $[n(\ell),2,d(\ell)]$ triangular code $\mathcal{T}_\ell$, we search over all logical operators and compute their weight.

We interpret the stabilisers as linear constraints over $\mathbb{F}_2$, so that $X$-type logical operators correspond to vectors $\vec{x} \in \ker H_\ell$, where $H_\ell$ is the parity-check matrix of $\mathcal{T}_\ell$, and $x_i = 1$ (resp.\ $x_i = 0$) indicates that the logical operator acts non-trivially (resp.\ trivially) on qubit $i$.

A logical operator represented by $\vec{x} \in \ker H_\ell$ must commute with all stabilisers.
As a result, weight-two stabilisers enforce equality $x_i = x_j$ between neighbouring qubits $i, j$ along each boundary.
Thus, all qubits on the left (resp.\ right) boundary share a common value, determined by the bottom-left (resp.\ bottom-right) qubit.
We denote these boundary values by $x_1$ and $x_2$, respectively.

Similarly, weight-three stabilisers impose that the parity of the three qubits on each triangular face vanishes.
Formally, for any triangle with vertices $i,j,k$, we have $x_k = x_i + x_j \mod 2$.

Under these constraints, a logical operator is fully specified by the two bottom qubits, represented by the two boundary variables $x_1$ and $x_2$.
Indeed, proceeding row by row from the bottom, each non-boundary qubit is the apex of a unique triangle whose other two vertices lie in the row below.
The commutation constraints with the stabilisers then fix the value of the remaining top qubit as the modulo-two sum of the two bottom qubits.
Since each qubit is determined exactly once in this process, this yields a unique global assignment.
All logical operators are consequently obtained by choosing $(x_1, x_2) \in \mathbb{F}_2^2$.
This gives
\begin{subequations}
\begin{align}
    &\id \id_L: (x_1,x_2) = (0,0), \\
    &X \id_L: (x_1,x_2) = (1,0), \\
    &\id X_L: (x_1,x_2) = (0,1), \\
    &XX_L: (x_1,x_2) = (1,1).
\end{align}
\end{subequations}

\begin{figure}
        \centering
        \includegraphics[width=0.65\linewidth]{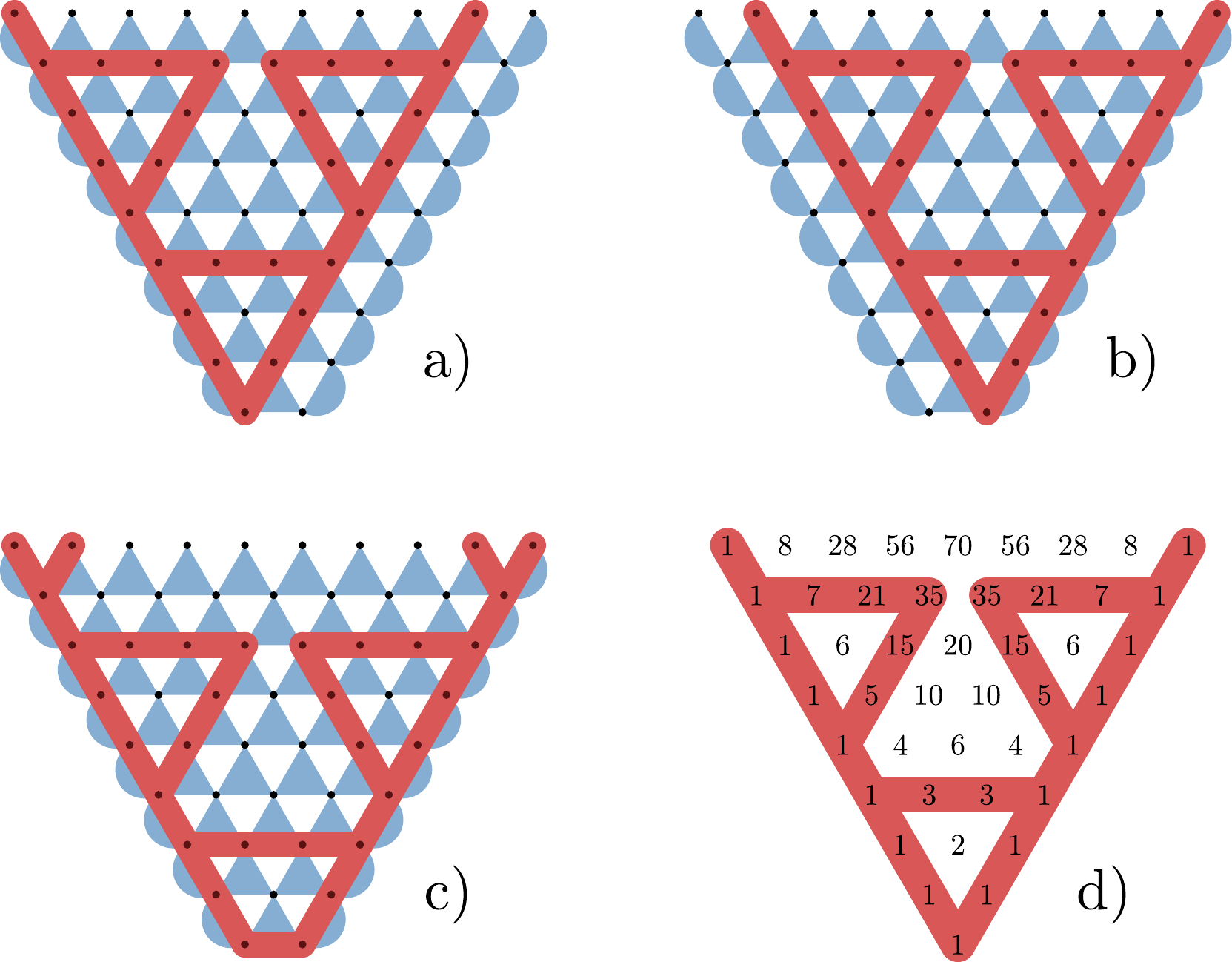}
        \caption{Logical operators of the triangular code $\mathcal{T}_8$, which encodes $k=2$ logical qubits. 
        The logical operators are $X$-type operators supported on the data qubits highlighted by the red regions. 
        Panels a), b), and c) show $X \id_L$, $\id X_L$, and $X X_L$, respectively.  
        Fixing the values of the two bottom qubits uniquely determines the remainder of the logical operator via addition modulo $2$, since it must commute with all stabilisers. 
        This construction reproduces Pascal’s triangle modulo $2$.
        Panel d) illustrates this connection by highlighting the odd entries of Pascal's triangle.}
        \label{fig:app:cell_auto_log_ops_construction}
    \end{figure}

This iterative construction is mathematically equivalent to generating Pascal’s triangle modulo $2$, which forms a fractal known as Sierpiński triangle \cite{wolfram1984geometry}. 
The resulting logical operators, whose support corresponds to the odd entries, therefore form a Sierpiński triangle, as shown in \cref{fig:app:cell_auto_log_ops_construction}.

This connection directly gives us the weights of the non-trivial logical operators.
Let $\{G_\ell\}_{\ell \in \mathbb{N}}$ denote the partial sums of Gould's sequence, which counts all odd numbers in the first $\ell+2$ rows of Pascal's triangle.
Then,
\begin{subequations}
    \begin{align}
        &\mathrm{wt}(X \id_L) = \mathrm{wt}(\id X_L) = G_{\ell+1}, \\
        &\mathrm{wt}(X X_L) = G_{\ell+2} - 1. 
    \end{align}
\end{subequations}
Since $G_{\ell+2} \geq G_{\ell+1} + 2$ for all $\ell \geq 1$, the distance of $\mathcal{T}_\ell$ is given by $d(\ell) = G_{\ell+1}$.

\subsection{Matching graph and syndrome extraction circuit}

\begin{figure}
        \centering
        \includegraphics[width=0.45\linewidth]{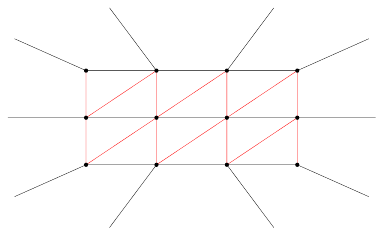}
        \caption{Matching graph of the triangular code $\mathcal{T}_1 = [5,2,3]$ under circuit-level noise, for $r=d$ rounds of syndrome extraction. 
        Vertices represent detectors and edges represent error mechanisms~\cite{gidney2021stim}. 
        Red edges indicate error mechanisms that also flip a logical observable.}
        \label{fig:matching_graph_cellular_automata_d3}
    \end{figure}

\begin{figure}
        \centering
        \includegraphics[width=0.75\linewidth]{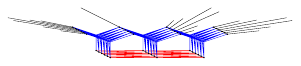}
        \caption{Matching graph of the triangular code $\mathcal{T}_2 = [9,2,5]$ under circuit-level noise, for $r=d$ rounds of syndrome extraction. 
        Vertices represent detectors and edges represent error mechanisms~\cite{gidney2021stim}. 
        Red edges indicate error mechanisms that also flip a logical observable. 
        Blue hyperedges indicate error mechanisms that flip multiple detectors.}
        \label{fig:matching_graph_cellular_automata_d5}
    \end{figure}

Figures \ref{fig:matching_graph_cellular_automata_d3} and \ref{fig:matching_graph_cellular_automata_d5} show the matching graphs of $\mathcal{T}_1$ and $\mathcal{T}_2$, respectively, under circuit-level noise.
The main difference is the presence of hyperedges in $\mathcal{T}_2$, which may cause BP+OSD to deviate from other decoders at low physical error rates.
We leave a detailed analysis of this hypothesis to future work.

\begin{figure}
        \centering
        \includegraphics[width=\linewidth]{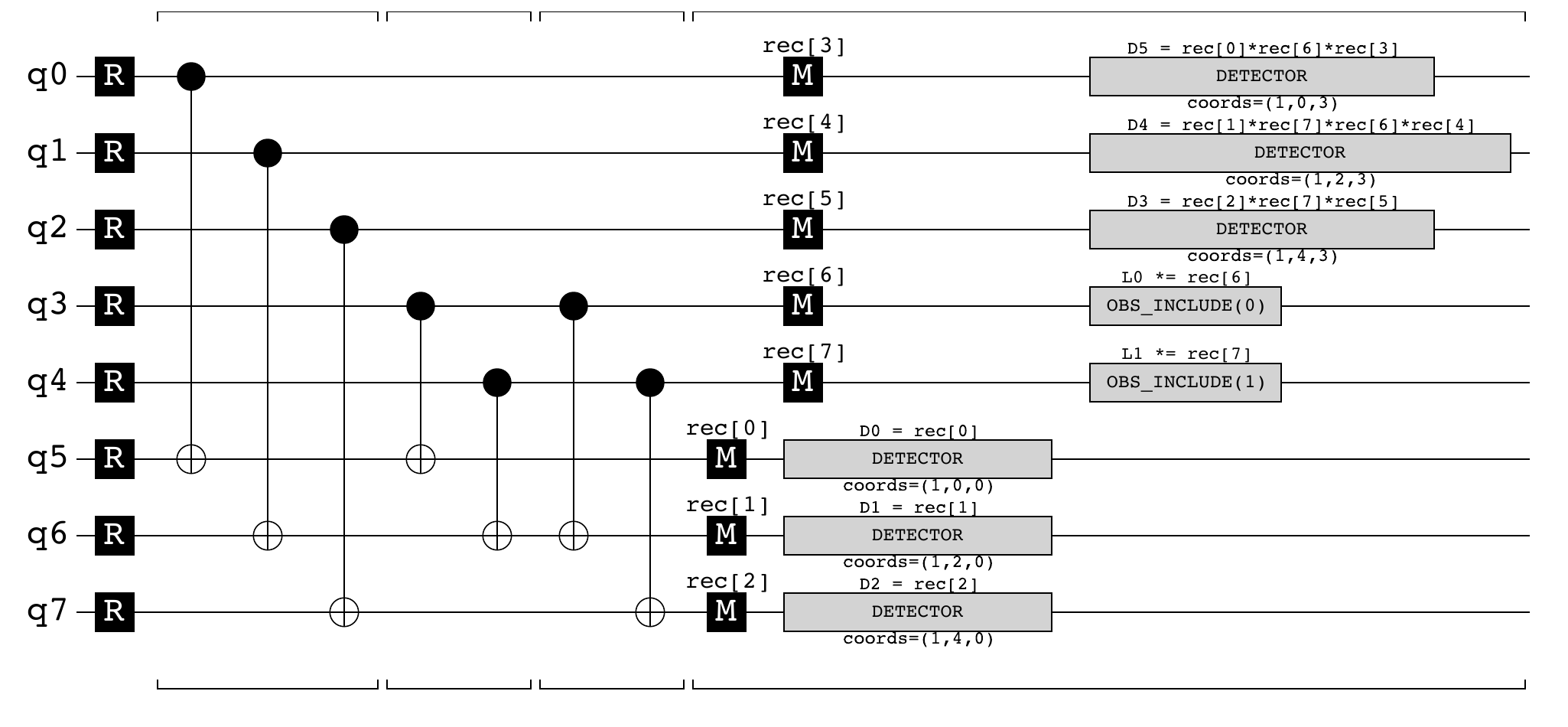}
        \caption{Noiseless circuit for $r=1$ round of syndrome extraction for the triangular code $\mathcal{T}_1=[5, 2, 3]$.}
        \label{fig:circuit_cellular_automata_d3}
    \end{figure}

\begin{figure}
        \centering
        \includegraphics[width=\linewidth]{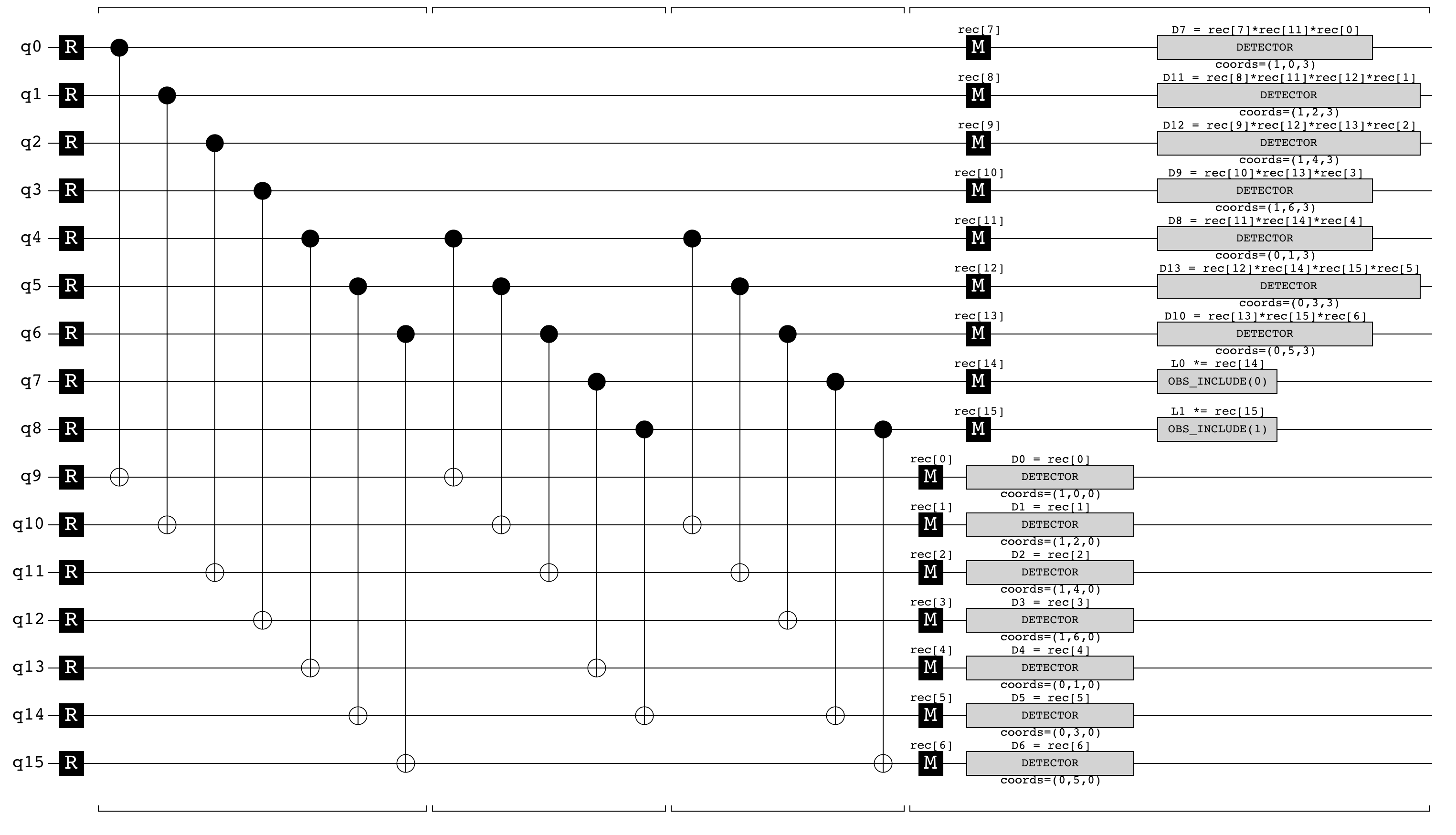}
        \caption{Noiseless circuit for $r=1$ round of syndrome extraction for the triangular code $\mathcal{T}_2 = [9, 2, 5]$.}
        \label{fig:circuit_cellular_automata_d5}
    \end{figure}

Figures \ref{fig:circuit_cellular_automata_d3} and \ref{fig:circuit_cellular_automata_d5} show the circuits for $r=1$ round of noiseless syndrome extraction for the triangular codes $\mathcal{T}_1$ and $\mathcal{T}_2$, respectively.

    
\end{document}